\documentclass[preprint,aps]{revtex4}

\usepackage{graphicx}
\usepackage{dcolumn}
\usepackage{bm}

\begin{document}

\preprint{PRL}

\title{Electron tunnel rates in a donor-silicon single electron transistor hybrid}

\author{Hans Huebl$^{1,2}$, Christopher D. Nugroho$^{1,3}$, Andrea Morello$^1$, Christopher C. Escott$^1$, Mark A. Eriksson$^4$, Changyi Yang$^5$, David N. Jamieson$^5$, Robert G. Clark$^1$, and Andrew S. Dzurak$^1$}
\email[corresponding author ]{huebl@wmi.badw.de}

\affiliation{$^1$ Australian Research Council Centre of Excellence for Quantum Computer Technology, Schools of Electrical Engineering and Physics, The University of New South Wales, Sydney NSW 2052, Australia\\
$^2$ Walther-Meissner-Institut, Bayerische Akademie der Wissenschaften, 85748~Garching, Germany\\
$^3$ Department of Physics, University of Illinois at Urbana-Champaign, Urbana, Illinois 61801, USA\\
$^4$ Department of Physics, University of Wisconsin, Madison, Wisconsin 53706, USA\\
$^5$ Australian Research Council Centre of Excellence for Quantum Computer Technology, School of Physics, The University of Melbourne, Melbourne VIC 3010, Australia}

\date{\today}

\begin{abstract}
We investigate a hybrid structure consisting of $20\pm4$ implanted
$^{31}$P atoms close to a gate-induced silicon single electron
transistor (SiSET). In this configuration, the SiSET is extremely
sensitive to the charge state of the nearby centers, turning from
the off state to the conducting state when the charge configuration
is changed. We present a method to measure fast electron tunnel
rates between donors and the SiSET island, using a pulsed voltage scheme and low-bandwidth
current detection. The experimental findings are quantitatively
discussed using a rate equation model, enabling the extraction of
the capture and emission rates.
\end{abstract}


\maketitle
The readout of a single spin is one of the key elements in
spin-based quantum information processing schemes
\cite{kane98,cerletti05}. One may distinguish between single-shot
readout, where the projective measurement of a single spin is
performed in real time, and ``spectroscopic'' readout, where the
expectation value of the spin state is deduced from a time-averaged
quantity (e.g. electrical current, fluorescence emission, \ldots).
Single-shot readout has been demonstrated in GaAs/AlGaAs quantum
dots \cite{elzerman04,barthel09}, while spectroscopic readout has
been obtained in a variety of systems, from quantum dots
\cite{hanson07,berezovsky06,shaji08, liu05b, xiao04} to NV centers in diamond
\cite{jelezko04} and dopant atoms in silicon \cite{sellier06}.
Carbon and silicon are particularly attractive platforms for
solid-state spin-based quantum processors, because they can be
isotopically purified to minimize decoherence induced by nuclear
spins. However, single-shot readout in these systems has not yet
been demonstrated. Recently, an architecture for single-shot readout
of a donor spin in silicon was proposed \cite{morello09},
consisting of a single implanted P donor \cite{jamieson05} in close
proximity to an induced silicon single-electron transistor (SiSET)
\cite{angus07}. The approach employs a readout principle similar to
the one successfully demonstrated in GaAs/AlGaAs single quantum dots
\cite{elzerman04}, where the spin state of the electron is deduced
from the time-resolved observation of spin-dependent tunneling
between the dot and a charge reservoir. However, in the donor-based
proposal \cite{morello09}, the bulk charge reservoir is replaced by
the island of a SiSET. This configuration is predicted to yield very
large charge transfer signals, thereby allowing high-fidelity
single-shot spin readout. The timescale of the projective spin
measurement is set by the electron tunneling time between donor and
SiSET, which must be controlled and understood before attempting
spin readout.

In this Letter we demonstrate and investigate the tunneling of
electrons in a hybrid device, consisting of approx.\ $20\pm4$
$^{31}$P donors, implanted next to an induced SiSET. We
show that the current through the SiSET, $I_{\rm SET}$, can be
switched from zero to the maximum value by transferring an electron
from a charge center to the SET island. By applying voltage pulses to a gate
near the donors while monitoring $I_{\rm SET}$, we study the
probability for an electron to tunnel between the center and the
SiSET. The resulting change in $I_{\rm SET}$ can be understood by
considering the donor-SET hybrid system as analogous to a double
quantum dot in the parallel configuration \cite{hofmann95}. We find
that the amplitudes in a pattern of Coulomb peaks depend on the pulse
duration and duty cycle, relative to the emission and capture rates
for tunneling from or onto the donor. Employing a rate equation
model, we are able to extract the electron tunneling rate for a
specific charge center. We observe fast tunneling rates of
3000~s$^{-1}$ for the tunnel event to load the center and
1000~s$^{-1}$ for the reverse process, despite the detection
bandwidth for the SET (DC) readout being limited to 200~Hz.

\begin{figure}
\includegraphics[width=6.5cm]{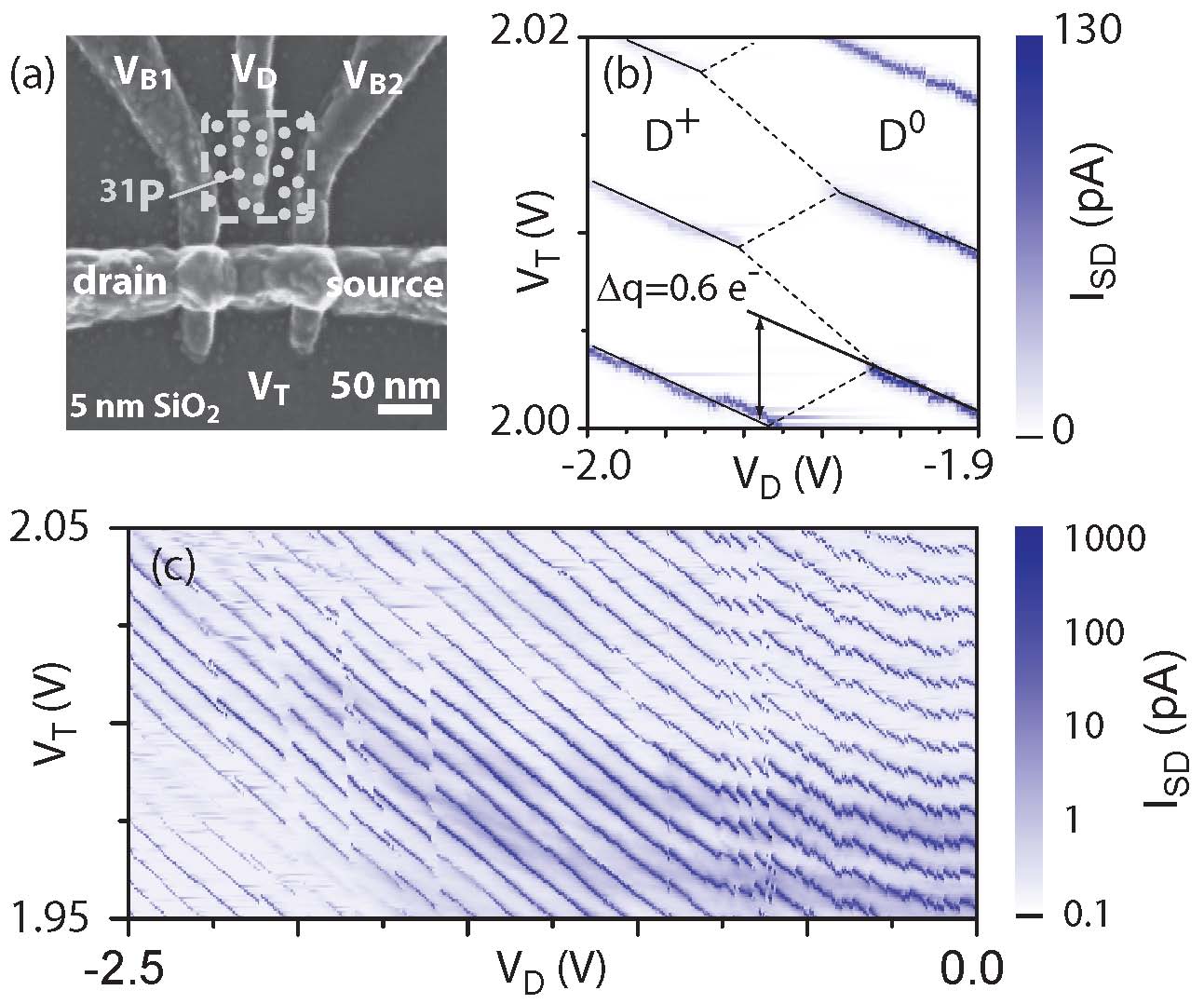}
\caption{Scanning electron micrograph of a hybrid device. The P
donors are implanted close to an induced SiSET (grey dashed square). The SiSET is formed by two gate
controlled ($V_{B1}$,$V_{B2}$) tunnel junctions and the overlapping
top-gate.  Panel (b) displays a close up of the stability
diagram near the charge transition of a center with large charge
transfer signal $\Delta q \sim 0.6e$. A large gate voltage scan is
shown in (c), where various charge transitions from multiple centers
are visible.}\label{fig:sample}
\end{figure}
Figure~\ref{fig:sample}~(a) shows the device fabricated on a high
purity intrinsic silicon wafer ($>10$~k$\Omega$cm), with the
implantation sites (grey dots) located next to an induced
SiSET~\cite{angus07,angus08}. In the active device region a
high-quality, 5~nm thick silicon oxide is grown by dry thermal
oxidation, yielding a very low density of interface traps $\sim
10^{10}$/eV/cm$^2$ near the conduction band edge \cite{johnson09}.
Underneath this oxide ohmic contacts are provided
([P]$=$5$\times 10^{19}$ cm$^{-3}$). In a first electron beam
lithography (EBL) step, with subsequent development and evaporation,
Ti(15~nm)/Pt(65~nm) alignment markers are formed for a high
precision ($< 20$~nm) realignment of subsequent layers. A
$90\times90~\rm{nm^2}$ aperture is opened in the PMMA resist, acting
as mask for the $^{31}$P donors, which are implanted with an acceleration
voltage of 14~keV and at a fluence of $2.5\times 10^{11}$~cm$^{-2}$,
resulting in a total of $20\pm 4$~$^{31}$P donors in this region.
After a rapid thermal anneal (1000~$^\circ$C, 5~s) to repair the
implantation damage, the Al donor control gate as well as the Al
barrier gates of the SiSET are patterned. The surface of these gates
is oxidized by an O$_2$ plasma ash for 4~min at 180$^\circ$C,
resulting in a $\sim 5$~nm thick Al$_x$O$_y$ insulating layer
\cite{lim09}. An Al top-gate, overlaying the barriers and the
source-drain regions, is formed in the last EBL step. This process
results in a hybrid quantum system with a few $^{31}$P donors in
close vicinity to a SiSET. The sample is operated in a dilution
refrigerator at an electron temperature $\approx 200$~mK. The
source, drain, as well as the SiSET control gates are connected to
the room temperature electronics via Cu powder filters with a cut
off frequency $\sim 1$~GHz. The donor control gate is connected via
a high-bandwidth line to apply high frequency pulses ($f_{{\rm 3dB}}\sim 500$~kHz, limited by resistive voltage dividers at room temperature). Its voltage
$V_{\rm D}$ is the sum of a constant component plus a rectangular
wave for pulsed voltage spectroscopy. The SiSET DC source-drain
current is measured using a current amplifier with 200~Hz bandwidth
and a gain of 10$^{10}$V/A.

\begin{figure}
\includegraphics[width=6cm]{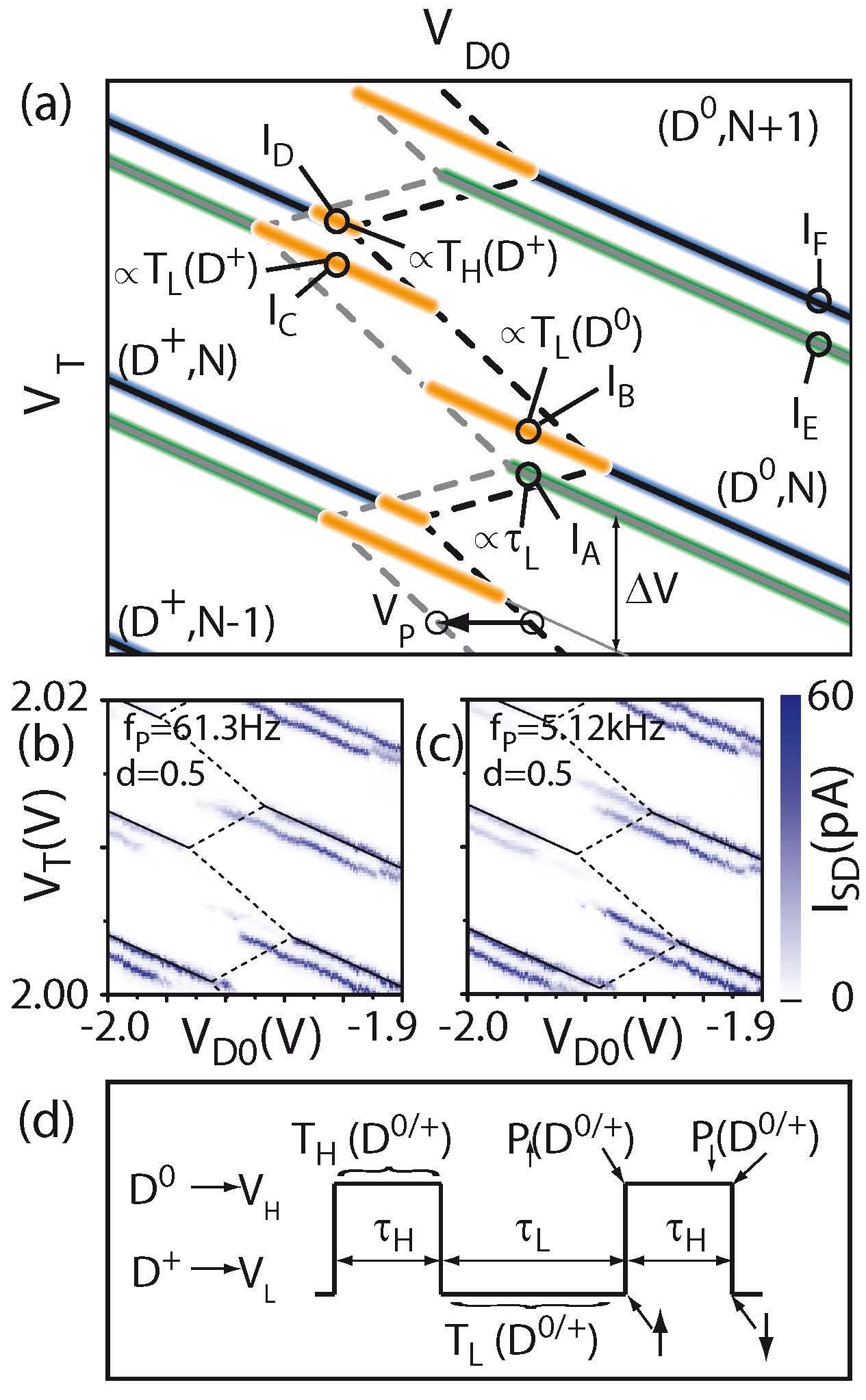}
\caption{Charge stability diagram for the pulsed voltage
spectroscopy. (a) Sketch of the SiSET conductance as function of the
top gate voltage $V_{\rm T}$ and the dc component of the donor control
gate $V_{\rm D0}$. Drawn as blue (green) lines are the positions of
the Coulomb peaks when the added pulse voltage $V_{\rm P}$=$V_{\rm L}$=0 V
($V_{\rm P}$=$V_{\rm H}$). The dashed lines are guides to the eye to indicate
a slice of the hexagonal charge stability diagram characteristic for
a double dot system.  When the frequency of the square wave $f_{\rm P}$
for switching between $V_{\rm L}$ and $V_{\rm H}$ is much smaller than the
tunneling rates $\Gamma_{\rm c}$ and $\Gamma_{\rm e}$, no current is observed
between the two superimposed stability diagrams (e.g. panel (b), where $f_{\rm P}=61.3$~Hz with duty cycle $d=0.5$),
because the charge configuration can follow the equilibrium state.
In contrast, when $2 \pi f_{\rm P}\gg \Gamma_{\rm c},\;\Gamma_{\rm e}$ a non-equilibrium
charge configuration can be observed resulting in current along the
orange lines. Panel (c) displays $I_{\rm SET}$ for
$f_{\rm P}=5.12$~kHz, where additional current is visible in this area.
Panel (d) shows the schematic for the pulsed voltage imposing a
charge transition, indicating the various relevant times for the
rate equation model as described in the
text.\label{fig:stability_diagram}}
\end{figure}
The capacitively- and tunnel-coupled donor and the SiSET island
effectively form a double quantum dot in parallel configuration
\cite{hofmann95}. Both series and parallel configurations result in
a hexagonal stability diagram, but in the series configuration,
transport only occurs at the triple points\cite{vanderwiel03}. In
contrast, in the parallel configuration the transport channel is
open for any gate voltages for which the electrochemical potential
of the SiSET, $\mu_{\rm SET}$, resides in the source-drain bias
window. As a result, transport occurs along some of the lines that
connect the triple points, which we call transport-lines in the following. Fig.~\ref{fig:sample}~(b) shows these
transport-lines in the vicinity of a charge transition, measured with a
source-drain bias $V_{\rm SD}$ $=$ 50 $\mu$V. The relevant gate space is
defined by the top-gate ($V_{\rm T}$) of the SiSET and the donor
control gate ($V_{\rm D}$). When the energy level of the donor is raised
with respect to $\mu_{\rm SET}$, at the charge transition point it
becomes favorable to remove an electron. This change in the charge
configuration (here labeled as ${\rm{D^0}} \rightarrow {\rm{D^+}}$
transition) acts back on $\mu_{\rm SET}$ and results in a shift of
the Coulomb peak lines. The magnitude of the shift in $\mu_{\rm
SET}$, relative to the Coulomb peak spacing, is quantified by the
charge transfer signal $\Delta q\approx0.6e$. Since $\Delta q$ is
much larger than the width of the Coulomb peaks, $I_{\rm SET}$ is
switched from zero to its maximum value by changing the occupancy
of the charge center. The data in Fig.~\ref{fig:sample}~(b)
demonstrate the ability to resolve with essentially 100\% contrast the charge
state of the donor, a critical prerequisite for the spin readout
method proposed in Ref.~\onlinecite{morello09}.

As shown in Fig~\ref{fig:sample}~(c), the measurement of $I_{\rm
SET}$ as a function of $V_{\rm T}$ and $V_{\rm D}$ yields a set of Coulomb
peaks appearing as tilted lines (due to the cross-capacitance
between control gate and SET island) that break at the charge
transition points. For $V_{\rm D}$ $>$ -$0.6$ V the slope of the transport-lines decreases, indicating charge accumulation under the donor control gate. In this regime we find several small charge transitions with $\Delta q < 0.1$, which we interpret as the ionization of shallow charge centers to the Si/SiO$_2$ interface. At more negative voltages, the pattern clears up, showing well-isolated charge transfers with 0.2 $<$ $\Delta q$ $<$ 0.6 in agreement with the predicted values for electrons tunneling into the SET island from donors $\sim 30 - 50$~nm away \cite{morello09}, and similar to the values observed in \cite{sun09} for a charge center near Al- and Si-SETs. This part of the stability diagram is stable and reproducible upon thermal cycling to room temperature.

We stress that the parallel geometrical
configuration of our hybrid device impedes direct spectroscopy of
the charge center coupled to the SiSET. It is therefore impossible
to distinguish a donor from e.g.~an interface trap on the basis of
its energy levels structure \cite{landsbergen08}. The number of
charge transitions observed for $V_{\rm D}$ $<$ -0.6 V is compatible
with the number of donors expected to be found within 30 to 50~nm
from the SET island, given the P implant fluence.
Furthermore, we note that the charge transitions in this regime typically group in pairs, agreeing in $\Delta q$ and slope in the $V_{\rm T}$-$V_{\rm D}$ gate space, again compatible with the observations of P donors with two charge transition levels expected \cite{sellier06, ramdas81}.
However, the unambiguous identification of the charge center remains a quest for
spin readout in combination with magnetic resonance techniques
\cite{morello09}. In this letter, the main focus is on the study of
tunnel rates between a SiSET and a charge center, whose
precise nature does not affect the results.

We measure the tunnel rates by superimposing on the DC voltage of
the donor control gate ($V_{\rm D0}$) a rectangular wave with
frequency $f_{\rm P}$, duty cycle $d$, and amplitude $V_{\rm P}$ (cf.~Fig.~\ref{fig:stability_diagram}~(d)). If
$f_{\rm P}$ is slow compared to the (de)charging rate of the
center, we record two stability diagrams (blue and green in
Fig.~\ref{fig:stability_diagram}~(a)), offset by $V_{\rm P}$ on the
horizontal axis when plotted vs.\ the DC gate voltages $V_{\rm D0}$
and $V_{\rm T}$. These arise because any point on the diagram probes
the average $I_{\rm SET}$ for the combination of the gate voltages
$(V_{\rm D0}+0~{\rm{V}},V_{\rm T})$ and $(V_{\rm D0}+V_P,V_{\rm T})$.
Conversely, if $f_{P}$ is faster than the electron tunnel rate
to/from the charge center, we find $I_{\rm SET} \neq 0$ at gate
configurations where transport would be otherwise suppressed, which we call non-equilibrium transport-lines in the following
(cf.\ orange lines in Fig.~\ref{fig:stability_diagram}), in addition to
the pure shifting of the pattern along the $V_{\rm D0}$-axis. These lines
arise because the charge center retains its configuration for
the time-span determined by the tunneling time, even while its chemical
potential crosses the charge transition point. To be specific, at
$I_B$ (cf.~Fig.~\ref{fig:stability_diagram}~(a)) no current is
expected for the $D^+$ configuration, the equilibrium state at $V_{\rm P}
= 0$. When the additional voltage pulse is in the high
state, the chemical potential is pushed over the charge transition
point, into the region where $D^0$ is the equilibrium configuration.
If an electron is captured (and $D^0$ is occupied)
during this time, immediately after $V_{\rm P}$ is brought back to zero we will find
the $D^0$ state at a gate configuration where a transport-line is
present. Thus, $I_{\rm SET} \neq 0$ when $V_{\rm P} = 0$, until the
electron tunnels out again. Observing a DC current in the orange
shaded area around $I_B$ indicates that the system is able to
maintain a non-equilibrium state for a time comparable to the pulse
duration, i.e. the tunnel rate is comparable to the pulsing
frequency $f_{\rm P}$. Similar arguments hold for $I_C$ and $I_D$,
whereas the bias line around $I_A$ is not altered, because the
voltage pulse does not cause a charge transition of the center. Note
that this symmetry break stems from the asymmetric pulsing
between 0 V and $V_P$. The low-frequency limit is shown in
Fig.~\ref{fig:stability_diagram}~(b), where $f_{\rm P}=61.3$~Hz, and
only a horizontally-shifted duplicate of the Coulomb peaks pattern
is observed. In contrast, the data in Fig.~\ref{fig:stability_diagram}~
(c) illustrate the high-frequency limit $f_{\rm P}=5.12$~kHz, where we
find $I_{\rm SET}\neq 0$ at the location of non-equilibrium transport lines.

Quantitatively, the DC value of $I_{\rm SET}$ at the non-equilibrium
current peaks can be understood within a rate equation
model. When we pulse the chemical potential of the charge center
over the charge transition level, the current state D$^{+/0}$ will
either persist, because the stable state is reached, or change to
the opposite state with the corresponding capture ($\Gamma_{\rm c}$) or
emission ($\Gamma_{\rm e}$) rate. The probability to find at the point
$\downarrow$ in Fig.~\ref{fig:stability_diagram}~(d) the
${\rm{D^+}}$ state occupied is
$P_\downarrow({\rm{D^+}})=P_\uparrow({\rm{D^+}})
{\rm{exp}}(-\Gamma_C \tau_H)$, because during $\tau_H$ the system
tends towards ${\rm{D^0}}$. Additionally,
$P_\downarrow({\rm{D^0}})=P_\uparrow({\rm{D^0}})+P_\uparrow({\rm{D^+}})(1-{\rm{exp}}(-\Gamma_{\rm c}
\tau_H))$, because $P_\downarrow({\rm{D^0}})$ is increased during
$\tau_H$. The same arguments hold for the inverse
direction yielding
$P_\uparrow({\rm{D^0}})=P_\downarrow({\rm{D^0}}){\rm{exp}}(-\Gamma_{\rm e}
\tau_L)$ and
$P_\uparrow({\rm{D^+}})=P_\downarrow({\rm{D^+}})+P_\downarrow({\rm{D^0}})(1-{\rm{exp}}(-\Gamma_{\rm e}
\tau_L))$.
To determine the four separate time durations, corresponding to occupation of either ${\rm{D^0}}$ or ${\rm{D^+}}$, for both values or $V_{\rm P}$, which we label $T_{L/H}({\rm{D^0}}/{\rm{D^+}})$) we express the probabilities $P_{\uparrow/\downarrow}({\rm{D^0}}/{\rm{D^+}})$ as a function of $\Gamma_{\rm e}$ and $\Gamma_{\rm c}$. We integrate these probabilities, including the time evolution of the occupation of the charge state over the pulse length $\tau_{L/H}$, to obtain the average time, finding ${\rm{D^0}}/{\rm{D^+}}$ during $V_{\rm P}=V_0/V_H$.
The result is the four times of interest:
$T_L({\rm{D^{0}}})=(1/\Gamma_{\rm e}) S$,
$T_L({\rm{D^{+}}})=\tau_L-(1/\Gamma_{\rm e}) S$,
$T_H({\rm{D^{+}}})=(1/\Gamma_{\rm c}) S$, and
$T_H({\rm{D^{0}}})=\tau_H-(1/\Gamma_{\rm c}) S$, where
$S=\frac{(1-{\rm{exp}}(-\Gamma_{\rm c} \tau_H))(1-{\rm{exp}}(-\Gamma_{\rm e}
\tau_L))}{1-{\rm{exp}}(-\Gamma_{\rm c} \tau_H){\rm{exp}}(-\Gamma_{\rm e}
\tau_L)}$.

\begin{figure}
\includegraphics[width=6cm]{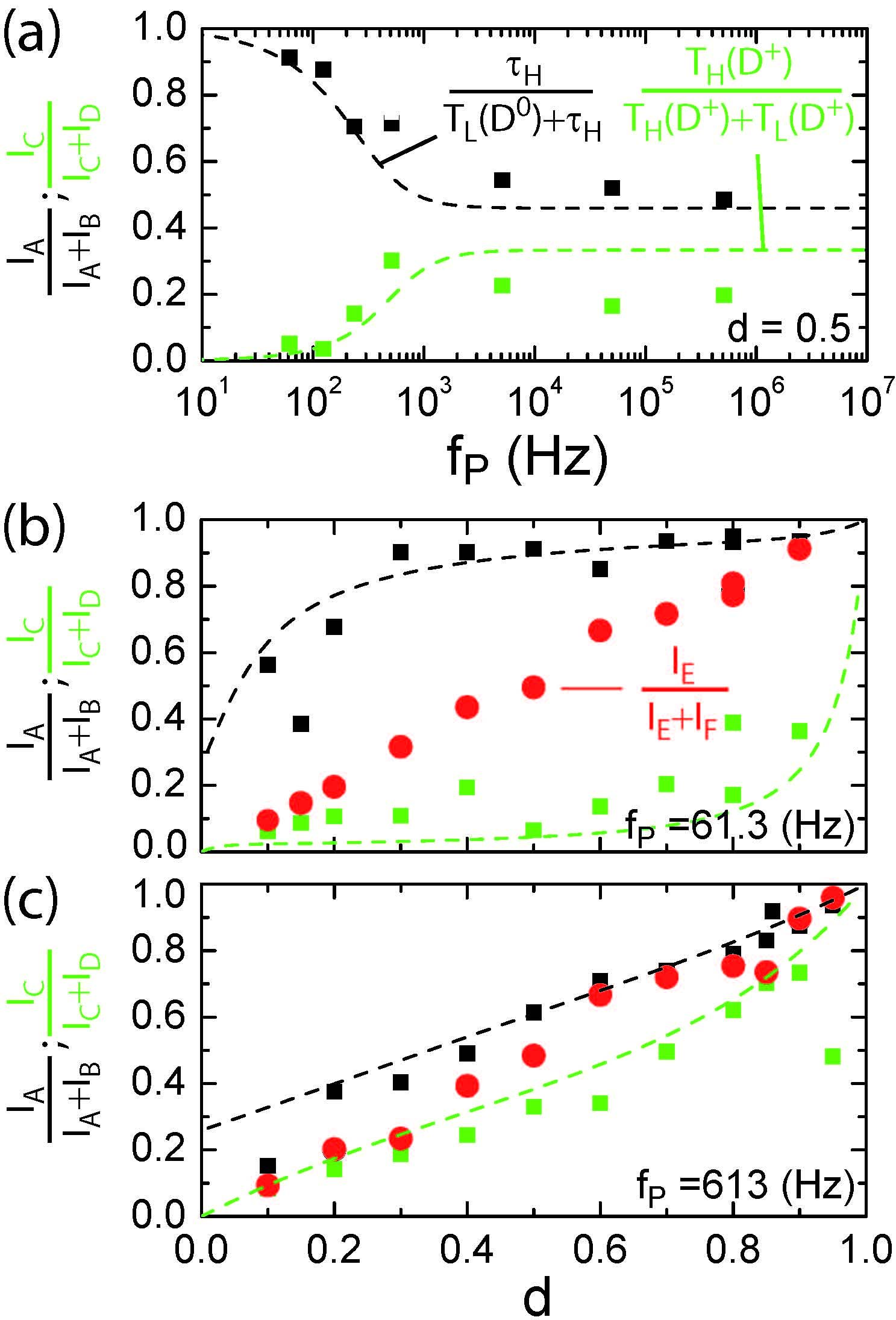}
\caption{Intensity ratios of the Coulomb peaks for
points in the charge stability diagram (cf. Fig.~\ref{fig:stability_diagram}~(a)) and time ratios originating
from the rate equation model (dashed lines). Panel (a) shows the
peak ratios as functions of the pulse frequency $f_{\rm P}$ for a
duty cycle of 0.5. The dashed lines represent the model using a
capture rate $\Gamma_{\rm c}=3000~{\rm{s^{-1}}}$ and an emission rate
$\Gamma_{\rm e}=1000~{\rm{s^{-1}}}$. Panels (b) and (c) show the same
ratios as function of the duty cycle for a fixed $f_{\rm P}$ of 61.3~Hz
and 613~Hz, respectively, where experiment and model is in good agreement. Additionally, the red filled circles
are the Coulomb peak ratios outside the region of the charge
transition. They show good agreement with the ratios of  $\tau_H$ and $\tau_L$  depending on the duty cicle $d$. \label{fig:tunnelrates}}
\end{figure}

The times obtained in this manner are proportional to $I_{\rm SET}$.
Since every transport-line has an individual current amplitude, we
analyze the peak ratios $\frac{I_B}{I_A+I_B}$
and
$\frac{I_D}{I_C+I_D}$
which are equal to  the ratios
$\frac{\tau_H}{\tau_H+T_L(D^0)}$
and
$\frac{T_H(D^+)}{T_H(D^+)+T_L(D^+)}$,
respectively. Although pulsing
is performed parallel to the control gate axis, it is possible to
compare current amplitudes from a  cut along the top-gate axis, because the current amplitude
does not vary significantly along specific transport-lines.
Figure~\ref{fig:tunnelrates}~(a) displays the peak ratios for a duty
cycle of $d=0.5$ as function of $f_{\rm P}$. The ratios
$\frac{I_B}{I_A+I_B}$ (black squares) and $\frac{I_D}{I_C+I_D}$
(green squares) are obtained from data like
Fig.~\ref{fig:stability_diagram}~(b) or (c) and both show a
quantitative agreement with the model using $\Gamma_{\rm e}=1000$~s$^{-1}$
and $\Gamma_{\rm c}=3000$~s$^{-1}$ over the entire frequency range
$f_P$. Fig.~\ref{fig:tunnelrates}~(b) compares the model
with the experimental data for a fixed $f_{\rm P}=61.3$~Hz as a function
of the duty cycle $d$. Again, the data is described
well by the model using the same capture and emission rates. At $d\approx0$ and $d\approx1$, the peak ratios are more difficult to determine due to the low $I_{\rm SET}$ for one of the contributions, explaining the deviations from the model.
For comparison, the duty cycle (red circles in
Fig.~\ref{fig:tunnelrates}~(b),(c)) is recovered from the spectra
independently by analyzing the ratio $\frac{I_E}{I_E+I_F}$,
showing good agreement with the duty cycle applied.
Fig.~\ref{fig:tunnelrates}~(c) shows the same plot as
Fig.~\ref{fig:tunnelrates}~(b) for a higher $f_{\rm P}=613$~Hz, again
in good agreement with the model.

An estimate of the distance between the charge center and the SET
island can be obtained from the capacitive modeling of the charge
transfer signal $\Delta q$, as shown in Ref.~\onlinecite{morello09}.
For the specific geometry of the device measured here, we find that
$\Delta q \sim 0.5e$ corresponds to a distance $\sim 40$~nm. We
use ISE-TCAD to calculate the profile of the conduction band
between donor and SET when the $D^0$ state is aligned with $\mu_{\rm
SET}$, and from this, the area of the tunnel barrier. A WKB
calculation of the tunnel rate
yields $\Gamma \sim 10^4$~s$^{-1}$, in reasonable
agreement with experimental findings.

In summary, we demonstrated and analyzed the tunneling of electrons
in a hybrid device consisting of $^{31}$P donors implanted next to
a gate-induced SiSET. We showed that the changes in the surrounding charge
configuration can be sensitively detected by the SET, and the mutual
coupling fulfills the requirements necessary for spin readout as
proposed in Ref.~\cite{morello09}. We further demonstrated a
technique to determine the tunnel rate of the center investigated,
and this technique is applicable even when this tunnel rate exceeds the
bandwidth of the detection SET. We
also provide a quantitative tunnel rate model that agrees with the
experimental findings. This experimental and theoretical toolbox
paves the way to the use of spin-dependent electron tunneling as a
readout method for single spins in silicon.

The authors thank D. Barber, N. Court, E. Gauja, R. P. Starrett, and K. Y. Tan for technical support at UNSW, and Alberto Cimmino for technical support at the Univ. of Melbourne. This work is supported by the Australian Research Council, the Australian
Government, and by the U.S. National Security Agency (NSA) and U.S.
Army Research Office (ARO) under Contract No. W911NF-08-1-0527. Work
at Wisconsin was supported by ARO and LPS under Contract No.
W911NF-08-1-0482.

\begin{thebibliography}{21}
\expandafter\ifx\csname natexlab\endcsname\relax\def\natexlab#1{#1}\fi
\expandafter\ifx\csname bibnamefont\endcsname\relax
  \def\bibnamefont#1{#1}\fi
\expandafter\ifx\csname bibfnamefont\endcsname\relax
  \def\bibfnamefont#1{#1}\fi
\expandafter\ifx\csname citenamefont\endcsname\relax
  \def\citenamefont#1{#1}\fi
\expandafter\ifx\csname url\endcsname\relax
  \def\url#1{\texttt{#1}}\fi
\expandafter\ifx\csname urlprefix\endcsname\relax\def\urlprefix{URL }\fi
\providecommand{\bibinfo}[2]{#2}
\providecommand{\eprint}[2][]{\url{#2}}

\bibitem[{\citenamefont{Kane}(1998)}]{kane98}
\bibinfo{author}{\bibfnamefont{B.~E.} \bibnamefont{Kane}},
  \bibinfo{journal}{Nature} \textbf{\bibinfo{volume}{393}},
  \bibinfo{pages}{133} (\bibinfo{year}{1998}).

\bibitem[{\citenamefont{Cerletti et~al.}(2005)\citenamefont{Cerletti, Coish,
  Gywat, and Loss}}]{cerletti05}
\bibinfo{author}{\bibfnamefont{V.}~\bibnamefont{Cerletti}} \bibnamefont{et al.},
  \bibinfo{journal}{Nanotechnology} \textbf{\bibinfo{volume}{16}},
  \bibinfo{pages}{R27} (\bibinfo{year}{2005}).

\bibitem[{\citenamefont{Elzerman et~al.}(2004)\citenamefont{Elzerman, Hanson,
  van Beveren, Witkamp, Vandersypen, and Kouwenhoven}}]{elzerman04}
\bibinfo{author}{\bibfnamefont{J.~M.} \bibnamefont{Elzerman}} \bibnamefont{et al.},
 \bibinfo{journal}{Nature}
  \textbf{\bibinfo{volume}{430}}, \bibinfo{pages}{431} (\bibinfo{year}{2004}).

\bibitem[{\citenamefont{Barthel et~al.}(2009)\citenamefont{Barthel, Reilly,
  Marcus, Hanson, and Gossard}}]{barthel09}
\bibinfo{author}{\bibfnamefont{C.}~\bibnamefont{Barthel}} \bibnamefont{et al.},
 \bibinfo{journal}{Phys. Rev. Lett.}
  \textbf{\bibinfo{volume}{103}}, \bibinfo{eid}{160503} (\bibinfo{year}{2009}).

\bibitem[{\citenamefont{Hanson et~al.}(2007)\citenamefont{Hanson, Kouwenhoven,
  Petta, Tarucha, and Vandersypen}}]{hanson07}
\bibinfo{author}{\bibfnamefont{R.}~\bibnamefont{Hanson}} \bibnamefont{et al.},
  \bibinfo{journal}{Rev. Mod. Phys.} \textbf{\bibinfo{volume}{79}},
  \bibinfo{pages}{1217} (\bibinfo{year}{2007}).

\bibitem[{\citenamefont{Berezovsky et~al.}(2006)\citenamefont{Berezovsky,
  Mikkelsen, Gywat, Stoltz, Coldren, and Awschalom}}]{berezovsky06}
\bibinfo{author}{\bibfnamefont{J.}~\bibnamefont{Berezovsky}} \bibnamefont{et al.},
 \bibinfo{journal}{Science}
  \textbf{\bibinfo{volume}{314}}, \bibinfo{pages}{1916} (\bibinfo{year}{2006}).

\bibitem[{\citenamefont{Shaji et~al.}(2008)\citenamefont{Shaji, Simons,
  Thalakulam, Klein, Qin, Luo, Savage, Lagally, Rimberg, Joynt
  et~al.}}]{shaji08}
\bibinfo{author}{\bibfnamefont{N.}~\bibnamefont{Shaji}} \bibnamefont{et al.},
 \bibinfo{journal}{Nature Physics}
  \textbf{\bibinfo{volume}{4}}, \bibinfo{pages}{540} (\bibinfo{year}{2008}).

\bibitem[{\citenamefont{Liu et~al.}(2005)\citenamefont{Liu, Fujisawa, Hayashi,
  and Hirayama}}]{liu05b}
\bibinfo{author}{\bibfnamefont{H.~W.} \bibnamefont{Liu}} \bibnamefont{et al.},
   \bibinfo{journal}{Phys. Rev. B} \textbf{\bibinfo{volume}{72}},
  \bibinfo{pages}{161305(R)} (\bibinfo{year}{2005}).

\bibitem[{\citenamefont{Xiao et~al.}(2004)\citenamefont{M. Xiao, I. Martin, E. Yablonovich and H. W. Jiang}}]{xiao04}
\bibinfo{author}{\bibfnamefont{M.} \bibnamefont{Xiao}} \bibnamefont{et al.},
   \bibinfo{journal}{Nature} \textbf{\bibinfo{volume}{430}},
  \bibinfo{pages}{435} (\bibinfo{year}{2004}).

\bibitem[{\citenamefont{Jelezko et~al.}(2004)\citenamefont{Jelezko, Gaebel,
  Popa, Gruber, and Wrachtrup}}]{jelezko04}
\bibinfo{author}{\bibfnamefont{F.}~\bibnamefont{Jelezko}} \bibnamefont{et al.},
    \bibinfo{journal}{Phys. Rev. Lett.} \textbf{\bibinfo{volume}{92}},
  \bibinfo{pages}{076401} (\bibinfo{year}{2004}).

\bibitem[{\citenamefont{Sellier et~al.}(2006)\citenamefont{Sellier, Lansbergen,
  Caro, Rogge, Collaert, Ferain, Jurczak, and Biesemans}}]{sellier06}
\bibinfo{author}{\bibfnamefont{H.}~\bibnamefont{Sellier}} \bibnamefont{et al.},
  \bibinfo{journal}{Phys. Rev. Lett.} \textbf{\bibinfo{volume}{97}},
  \bibinfo{pages}{206805} (\bibinfo{year}{2006}).

\bibitem[{\citenamefont{Morello et~al.}(2009)\citenamefont{Morello, Escott,
  Huebl, van Beveren, Hollenberg, Jamieson, Dzurak, and Clark}}]{morello09}
\bibinfo{author}{\bibfnamefont{A.}~\bibnamefont{Morello}} \bibnamefont{et al.},
  \bibinfo{journal}{Phys. Rev. B} \textbf{\bibinfo{volume}{80}},
  \bibinfo{pages}{081307(R)} (\bibinfo{year}{2009}).

\bibitem[{\citenamefont{Jamieson et~al.}(2005)\citenamefont{Jamieson, Yang,
  Hopf, Hearne, Pakes, Prawer, Mitic, Gauja, Andresen, Hudson
  et~al.}}]{jamieson05}
\bibinfo{author}{\bibfnamefont{D.~N.} \bibnamefont{Jamieson}} \bibnamefont{et al.},
\bibinfo{journal}{Appl. Phys. Lett.}
  \textbf{\bibinfo{volume}{86}}, \bibinfo{pages}{202101}
  (\bibinfo{year}{2005}).

\bibitem[{\citenamefont{Angus et~al.}(2007)\citenamefont{Angus, Ferguson,
  Dzurak, and Clark}}]{angus07}
\bibinfo{author}{\bibfnamefont{S.~J.} \bibnamefont{Angus}} \bibnamefont{et al.},
  \bibinfo{journal}{Nano Lett.} \textbf{\bibinfo{volume}{7}},
  \bibinfo{pages}{2051} (\bibinfo{year}{2007}).

\bibitem[{\citenamefont{Hofmann et~al.}(1995)\citenamefont{Hofmann, Heinzel,
  Wharam, Kotthaus, B\"{o}hm, Klein, Tr\"{a}nkle, and Weimann}}]{hofmann95}
\bibinfo{author}{\bibfnamefont{F.}~\bibnamefont{Hofmann}} \bibnamefont{et al.},
  \bibinfo{journal}{Phys. Rev. B} \textbf{\bibinfo{volume}{51}},
  \bibinfo{pages}{13872} (\bibinfo{year}{1995}).

\bibitem[{\citenamefont{Angus et~al.}(2008)\citenamefont{Angus, Ferguson,
  Dzurak, and Clark}}]{angus08}
\bibinfo{author}{\bibfnamefont{S.~J.} \bibnamefont{Angus}} \bibnamefont{et al.},
    \bibinfo{journal}{Appl. Phys. Lett.} \textbf{\bibinfo{volume}{92}},
  \bibinfo{pages}{112103} (\bibinfo{year}{2008}).

\bibitem[{\citenamefont{Johnson et~al.}(2009)\citenamefont{Johnson, McCallum,
  van Beveren, and Gauja}}]{johnson09}
\bibinfo{author}{\bibfnamefont{B.~C.} \bibnamefont{Johnson}} \bibnamefont{et al.},
   \bibinfo{journal}{Thin Solid Films, doi:10.1016/j.tsf.2009.09.152}  (\bibinfo{year}{2009}).

\bibitem[{\citenamefont{Lim et~al.}(2009)\citenamefont{Lim, Huebl, van Beveren,
  Rubanov, Spizzirri, Angus, Clark, and Dzurak}}]{lim09}
\bibinfo{author}{\bibfnamefont{W.~H.} \bibnamefont{Lim}} \bibnamefont{et al.},
   \bibinfo{journal}{Appl. Phys. Lett.} \textbf{\bibinfo{volume}{94}},
  \bibinfo{pages}{173502} (\bibinfo{year}{2009}).

\bibitem[{\citenamefont{van~der Wiel et~al.}(2003)\citenamefont{van~der Wiel,
  Franceschi, Elzerman, Fujisawa, Tarucha, and Kouwenhoven}}]{vanderwiel03}
\bibinfo{author}{\bibfnamefont{W.~G.} \bibnamefont{van~der Wiel}} \bibnamefont{et al.},
  \bibinfo{journal}{Rev. Mod. Phys.} \textbf{\bibinfo{volume}{75}},
  \bibinfo{pages}{1} (\bibinfo{year}{2003}).

\bibitem[{\citenamefont{Sun and Kane}(2009)}]{sun09}
\bibinfo{author}{\bibfnamefont{L.}~\bibnamefont{Sun}} \bibnamefont{and}
  \bibinfo{author}{\bibfnamefont{B.~E.} \bibnamefont{Kane}},
  \bibinfo{journal}{Phys. Rev. B} \textbf{\bibinfo{volume}{80}},
  \bibinfo{pages}{153310} (\bibinfo{year}{2009}).

\bibitem[{\citenamefont{Lansbergen et~al.}(2008)\citenamefont{Lansbergen,
  Rahman, Wellard, Woo, Caro, Collaert, Biesemans, Klimeck, Hollenberg, and
  Rogge}}]{landsbergen08}
\bibinfo{author}{\bibfnamefont{G.~P.} \bibnamefont{Lansbergen}} \bibnamefont{et al.},
   \bibinfo{journal}{Nature Physics} \textbf{\bibinfo{volume}{4}},
  \bibinfo{pages}{656} (\bibinfo{year}{2008}).

\bibitem[{\citenamefont{Ramdas and Rodriguez}(1981)}]{ramdas81}
\bibinfo{author}{\bibfnamefont{A.~K.} \bibnamefont{Ramdas}} \bibnamefont{and}
  \bibinfo{author}{\bibfnamefont{S.}~\bibnamefont{Rodriguez}},
  \bibinfo{journal}{Reports on Progress in Physics}
  \textbf{\bibinfo{volume}{44}}, \bibinfo{pages}{1297} (\bibinfo{year}{1981}).

\end{thebibliography}

\clearpage
\end{document}